\documentclass[conference]{IEEEtran}

\usepackage{etex}
\usepackage{amsmath}                        
\usepackage{booktabs}                       
\usepackage{graphicx}
\usepackage{amssymb, subfigure, stfloats, amssymb, amsfonts, rotating, acronym}
\usepackage{multirow}
\usepackage{relsize}
\usepackage{array}
\usepackage{tabularx}
\usepackage{tikz}
\usetikzlibrary{patterns}
\usetikzlibrary{arrows,positioning,shapes.geometric,spy}
\usepackage{url}
\usepackage{breakurl}
\usepackage{filecontents}
\usepackage{color}
\usepackage{xcolor}
\usepackage{pgfplots}
\usepackage{cite}
\usepackage{adjustbox}
\usepackage{colortbl}
\usepackage{lipsum}
\usepackage[linesnumbered,algoruled,boxed,lined]{algorithm2e}
\usepackage[normalem]{ulem}

\DeclareMathOperator*{\sgn}{sgn}

\newcommand{\fixme}[2]{\ifx&#2&{\leavevmode\color{red}#1}\else{\leavevmode\color{red}FIXME\{}#1{\leavevmode\color{red}\}}\footnote{{\leavevmode\color{red}#2}}\PackageWarning{Fixme}{#1: #2}\fi}

\newcommand{\newstuff}[2]{\ifx&#2&{\leavevmode\color{blue}#1}\else{\leavevmode\color{blue}NEWSTUFF\{}#1{\leavevmode\color{blue}\}}\footnote{{\leavevmode\color{blue}#2}}\PackageWarning{Newstuff}{#1: #2}\fi}

\newcommand{\furkan}[2]{\ifx&#2&{\leavevmode\color{magenta}#1}\else{\leavevmode\color{magenta}FIXME\{}#1{\leavevmode\color{magenta}\}}\footnote{{\leavevmode\color{magenta}#2}}\PackageWarning{Furkan}{#1: #2}\fi}

\definecolor{bblue}{HTML}{4F81BD}
\definecolor{rred}{HTML}{C0504D}
\definecolor{ggreen}{HTML}{9BBB59}
\definecolor{ppurple}{HTML}{9F4C7C}

\begin{document}

\title{Improved Successive Cancellation Flip Decoding of Polar Codes Based on Error Distribution\footnotemark}

\author{\IEEEauthorblockN{Carlo Condo, Furkan Ercan, Warren J. Gross}
\IEEEauthorblockA{Department of Electrical and Computer Engineering, McGill University, Montr\'eal, Qu\'ebec, Canada\\
Email: carlo.condo@mcgill.ca, furkan.ercan@mail.mcgill.ca, warren.gross@mcgill.ca}}

\maketitle

\begin{abstract}
Polar codes are a class of linear block codes that provably achieves channel capacity, and have been selected as a coding scheme for $5^{\rm th}$ generation wireless communication standards. Successive-cancellation (SC) decoding of polar codes has mediocre error-correction performance on short to moderate codeword lengths: the SC-Flip decoding algorithm is one of the solutions that have been proposed to overcome this issue. On the other hand, SC-Flip has a higher implementation complexity compared to SC due to the required log-likelihood ratio (LLR) selection and sorting process. Moreover, it requires a high number of iterations to reach good error-correction performance. In this work, we propose two techniques to improve the SC-Flip decoding algorithm for low-rate codes, based on the observation of channel-induced error distributions. The first one is a fixed index selection (FIS) scheme to avoid the substantial implementation cost of LLR selection and sorting with no cost on error-correction performance. The second is an enhanced index selection (EIS) criterion to improve the error-correction performance of SC-Flip decoding. A reduction of $24.6\%$ in the implementation cost of logic elements is estimated with the FIS approach, while simulation results show that EIS leads to an improvement on error-correction performance improvement up to $0.42$ dB at a target FER of $10^{-4}$.
\end{abstract}

\IEEEpeerreviewmaketitle

\section{Introduction}\label{sec:intro}

\IEEEPARstart{P}{olar} codes \cite{arikan09} are error-correcting codes proven to achieve channel capacity with infinite code length. They have been selected as a coding scheme in the $5^{\rm th}$ generation wireless systems standards (5G), within the enhanced mobile broadband communication scenario (eMBB), and are being considered for both ultra-reliable low-latency communications (URLLC) and massive machine-type communications (mMTC) \cite{3GPP-5G}. The variety of system parameters foreseen in 5G demand improved decoding algorithms with a focus on reliable communications and low power consumption.

\footnotetext{This version of the manuscript corrects an error in the previous ArXiv version, as well as the published version in IEEE Xplore under the same title, which has the DOI:10.1109/WCNCW.2018.8368991. The corrections include all the simulations of SC-Flip-based and SC-Oracle decoders, along with associated comments in-text.}

In \cite{arikan09}, the successive-cancellation (SC) decoding of polar codes was proposed. It is able to achieve channel capacity when the code length tends to infinite, but its error-correction performance degrades at shorter lengths. To overcome this limitation, the SC-List decoding algorithm was introduced in \cite{TalList}. The improved error-correction performance comes at the cost of longer latency and higher complexity: various contributions have improved speed and reduced implementation cost \cite{Alex_List_LLR,fastSSCL-jour,PSCL-JETCAS}.

Successive-cancellation flip (SC-Flip) decoding \cite{SCFlip14} takes a different approach to the improvement of SC, relying on an implementation complexity lower than that of SC-List, mainly sacrificing average decoding latency, and providing error-correction performance comparable to that of SC-List with a list size  of $2$.
SC-Flip relies on multiple subsequent applications of SC, and on the identification of the most likely points in which the SC algorithm made a wrong decision. This operation is based on the magnitude of logarithmic likelihood ratios (LLRs) associated to the estimated codeword bits, and it  is costly to implement. Modifications proposed in \cite{SCFlip17-conf,SCFlip17-jour} have shown that the error identification and correction process of SC-Flip can be greatly improved. A decoding approach similar to \cite{fastSSCL-jour} has been recently proposed for SC-Flip in \cite{giard_fastSCF}, reducing latency at no cost in error-correction performance, but substantially increasing the implementation complexity of the decoder.

In this work, we propose two techniques to improve SC-Flip decoding, based on the distribution of the first wrong estimation incurred by SC. The first method simplifies the wrong estimation identification process without degrading the error-correction performance, leading to substantial complexity reduction, while the second method uses the error distribution to restrict the search space to the bit indices most likely to incur errors. Simulation results show the effectiveness of the proposed techniques for low-rate polar codes, over a wide range frame error rate (FER) intervals.

The remainder of this work is organized as follows: in Section~\ref{sec:bg} polar codes and their decoding algorithms are introduced. In Section~\ref{sec:ISCFlip} the proposed techniques are detailed, and simulation results, together with comparison with the state of the art, are presented in Section~\ref{sec:sims}. Conclusions are drawn in Section~\ref{sec:concl}.

\section{Preliminaries}\label{sec:bg}
\subsection{Polar Codes}

A polar code $PC(N,K)$ of code length $N$ and rate $R=K/N$ is a linear block code that divides $N = 2^n, n \in \mathbb Z^+$ bit-channels in $K$ reliable ones and $N-K$ unreliable ones. Information bits are transmitted through the reliable channels, whereas unreliable channels are fixed to a value known by both transmitter and receiver, and are thus called frozen bits.

Polar codes are encoded through the following matrix multiplication:
\begin{equation}\label{eqn:enc}
\boldsymbol{x_0^{N-1}} = \boldsymbol{u_0^{N-1}}G^{\otimes n}\text{,}
\end{equation}
where $\boldsymbol{x_0^{N-1}} = \{x_0,x_1,\ldots,x_{N-1}\}$ represents the encoded vector, $\boldsymbol{u_0^{N-1}} = \{u_0,u_1,\ldots,u_{N-1}\}$ is the input vector, and the generator matrix $G^{\otimes n}$ is obtained as the $n$-th Kronecker product of the polarization matrix $G = \left[\begin{smallmatrix} 1&0\\ 1&1 \end{smallmatrix} \right]$. Due to the recursive nature of the encoding process, an $N$-length polar code can be interpreted as the concatenation of two polar codes of length $N/2$. 
%

\begin{figure}
  \centering
   \scalebox{0.75}{\begin{tikzpicture}[scale=.65, thick]

\draw [thin,gray,dashed] (0,-1) -- (7.5,-1);
\draw [thin,gray,dashed] (0,-3) -- (11.5,-3);
\draw [thin,gray,dashed] (0,-5) -- (13.5,-5);
\draw [thin,gray,dashed] (0,-7) -- (14.5,-7);

\node at (-.75,-1) {$S=3$};
\node at (-.75,-3) {$S=2$};
\node at (-.75,-5) {$S=1$};
\node at (-.75,-7) {$S=0$};

\draw (7.5,-1) -- (3.5,-3);
\draw (7.5,-1) -- (11.5,-3);

\draw (3.5,-3) -- (1.5,-5);
\draw (3.5,-3) -- (5.5,-5);
\draw (11.5,-3) -- (9.5,-5);
\draw (11.5,-3) -- (13.5,-5);

\draw (1.5,-5) -- (0.5,-7);
\draw (1.5,-5) -- (2.5,-7);

\draw (5.5,-5) -- (4.5,-7);
\draw (5.5,-5) -- (6.5,-7);

\draw (9.5,-5) -- (8.5,-7);
\draw (9.5,-5) -- (10.5,-7);

\draw (13.5,-5) -- (12.5,-7);
\draw (13.5,-5) -- (14.5,-7);


  \draw[black,fill=white] (.5,-7) circle [radius=.25];			
  \fill[color=gray] (1.5,-5) circle [radius=.25];	
  \draw[black,fill=white] (2.5,-7) circle [radius=.25];			
  \fill[color=gray] (3.5,-3) circle [radius=.25];	
  \draw[black,fill=white] (4.5,-7) circle [radius=.25];			
  \fill[color=gray] (5.5,-5) circle [radius=.25];	
  \fill[color=black] (6.5,-7) circle [radius=.25];	

  \fill[color=gray] (7.5,-1) circle [radius=.25];	

  \fill[color=black] (8.5,-7) circle [radius=.25];		
  \fill[color=gray] (9.5,-5) circle [radius=.25];		
  \fill[color=black] (10.5,-7) circle [radius=.25];		
  \fill[color=gray] (11.5,-3) circle [radius=.25];		
  \fill[color=black] (12.5,-7) circle [radius=.25];		
  \fill[color=gray] (13.5,-5) circle [radius=.25];		
  \fill[color=black] (14.5,-7) circle [radius=.25];		

\node at (.5,-7.8) {$\hat{u}_0$};
\node at (2.5,-7.8) {$\hat{u}_1$};
\node at (4.5,-7.8) {$\hat{u}_2$};
\node at (6.5,-7.8) {$\hat{u}_3$};
\node at (8.5,-7.8) {$\hat{u}_4$};
\node at (10.5,-7.8) {$\hat{u}_5$};
\node at (12.5,-7.8) {$\hat{u}_6$};
\node at (14.5,-7.8) {$\hat{u}_7$};

\draw [->] (8,-1.125) -- (11,-2.625) node [above=.05cm,midway,rotate=-25] {$\boldsymbol{\alpha}$};
\draw [<-] (8,-1.375) -- (11,-2.875) node [below=-.05cm,midway,rotate=-25] {$\boldsymbol{\beta}$};

\draw [->] (11.25,-3) -- (9.75,-4.5) node [above=.03cm,midway,rotate=40] {$\boldsymbol{{\alpha}^l}$};
\draw [<-] (11.25,-3.5) -- (9.75,-5) node [below=-.05cm,midway,rotate=40] {$\boldsymbol{{\beta}^l}$};

\draw [->] (11.75,-3) -- (13.25,-4.5) node [above=.03cm,midway,rotate=-40] {$\boldsymbol{{\alpha}^r}$};
\draw [<-] (11.75,-3.5) -- (13.25,-5) node [below=-.05cm,midway,rotate=-40] {$\boldsymbol{{\beta}^r}$};

\end{tikzpicture}}
  \\
  \vspace{2pt}
  \caption{Successive-cancellation decoding tree for a $PC(8,5)$ code.}
  \label{fig:scdecode}
\end{figure}
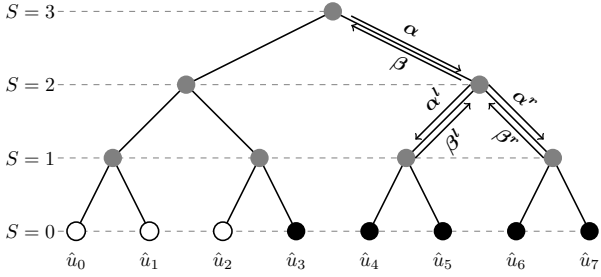

The scheduling of operations required by the SC decoding algorithm allows to see its process as a binary tree search, where the  tree is explored depth-first, with priority given to the left branch. 
Fig. \ref{fig:scdecode} portrays an example of SC decoding tree, for $PC(8,5)$. Each node receives from its parent a vector of LLRs $\boldsymbol{\alpha}=\{\alpha_0, \alpha_1, \ldots, \alpha_{2^S-1}\}$. Each node at stage $S$ computes the left $\boldsymbol{\alpha^l} = \{\alpha^l_0, \alpha^l_1, \ldots, \alpha^l_{2^{S-1}-1}\}$ and right $\boldsymbol{\alpha^r} = \{\alpha^r_0, \alpha^r_1, \ldots, \alpha^r_{2^{S-1}-1}\}$ LLR vectors sent to child nodes as
\begin{align}
{\alpha}^l_i &= \sgn(\alpha_{i})\sgn(\alpha_{i+2^{S-1}}) \min(|\alpha_{i}|,|\alpha_{i+2^{S-1}}|) \text{,} \label{eqn:alphaleft}\\
{\alpha}^r_i &= \alpha_{i+2^{S-1}} + (1-2\beta^{l}_{i})\alpha_{i} \text{.} \label{eqn:alpharight}
\end{align}
The LLRs at the root node are initialized as the channel LLR $\boldsymbol{y}_0^{N-1}$. Nodes receive the partial sums $\boldsymbol{\beta}$ from their left $\boldsymbol{\beta^{l}}=\{\beta^l_0, \beta^l_1, \ldots, \beta^l_{2^{S-1}-1}\}$ and right $\boldsymbol{\beta^{r}}=\{\beta^r_0, \beta^r_1, \ldots, \beta^r_{2^{S-1}-1}\}$ child node:
\begin{equation}\label{eqn:beta}
  \beta_i=\left\{
  \begin{array}{@{}ll@{}}
    \beta^{l}_{i} \oplus \beta^{r}_{i}, & \text{if}~ i \leq 2^{S-1} \\
    \beta^{r}_{i}, & \text{otherwise.}
  \end{array}\right.
\end{equation}
where $\boldsymbol{\oplus}$ is the bitwise XOR operation, and $0 \leq i < 2^S$. At leaf nodes, the $\beta$ value and the estimated bit vector $\boldsymbol{\hat{u}_0^{N-1}}$ are computed as
\begin{equation}\label{eqn:bitestimate-sc}
\beta_{i}=\left\{
  \begin{array}{@{}ll@{}}
    0, & \text{when } \alpha_i \geq 0 \text{ } \text{or } i \in \Phi; \\
    1, & \text{otherwise.}
  \end{array}\right.
\end{equation}

The SC-List decoding algorithm \cite{TalList} improves the error-correction performance of SC by relying on $L$ parallel SC decoding paths. Each path is associated to a metric, that helps deciding which of the $L$ paths is more likely to be correct. 

\subsection{Successive-Cancellation Flip Decoding}

To improve the error-correction performance of SC, in \cite{SCFlip14} the SC-Flip decoding algorithm was proposed. It is based on the observation that a failed SC decoding process is due to either an incorrect bit estimation caused by noise introduced by the channel or, due to the sequential nature of SC, to a previous incorrect bit estimation. Thus, the first wrong estimation is always caused by the channel noise. To evaluate the impact of wrong estimations, a genie-like decoder called SC-Oracle was created: using foreknowledge of the transmitted codeword, it identifies the channel-induced errors and ensures that SC estimates the bit correctly. SC-Oracle shows that avoiding all wrong decisions caused by the channel noise results in significantly improved SC error-correction performance. Moreover, simulations with SC-Oracle have shown that a failed SC decoding is mostly due to a single wrong decision caused by channel noise, an event that denote as $E_1$.

The SC-Flip algorithm attempts the identification and correction of $E_1$ events. A cyclic redundancy check (CRC) code with a $C$-bit remainder is used to encode the information bits. If the CRC check passes at the end of SC decoding, the estimated codeword is assumed correct. In case the CRC check detects an error, the $T_{max}$ LLRs with the smallest magnitude, representing the bit estimations with lowest reliability, are stored and sorted. The bit associated to the smallest LLR is flipped, and SC is is applied to the part of the decoding tree that follows it. This process is repeated for all the $T_{max}$ indices, or until the CRC passes.


In general, increasing $T_{max}$ improves the error-correction performance of SC-Flip, bringing it ever closer to that of SC-Oracle. In fact, the performance gap between SC-Flip and SC-Oracle is due to two possible cases: either the estimated codeword with a correct CRC check still contains errors, or the decoding stopped after reaching the maximum number of iterations $T_{max}$ without being successful. 

%

Further improvements for SC-Flip decoding algorithm have been recently proposed in \cite{SCFlip17-conf} and \cite{SCFlip17-jour}, where a generalized SC-Flip algorithm uses nested flips to correct more than one erroneous decision with a single CRC, with a simulation-based scaling metric also introduced to help the baseline SC-Flip decoder detect the erroneous bit indices more accurately. Simulation results show an improvement of $0.4$~dB when $T_{max}=10$ with respect to SC-Flip. The implementation of this method requires parallel decoders similar to SC-List.

\section{Improved Successive-Cancellation Flip Decoding}\label{sec:ISCFlip}

We propose two new techniques to improve the error-correction performance of SC-flip decoding, and to reduce the implementation complexity of SC-Flip decoders.

\subsection{Fixed Index Selection} \label{subsec:fixed}
In \cite{SCFlip14}, it was shown that the average number of iterations for SC-Flip decoding converges to that of a single SC decoder at moderate to high $E_b/N_0$. However, a single iteration of SC-Flip requires additional operations with respect to SC. The standard SC-Flip decoding algorithm decides which bits to flip by identifying the $T_{max}$ least reliable bit estimations among the non-frozen indices. Since the estimation reliability is directly associated to the magnitude of its LLR value (\ref{eqn:bitestimate-sc}), they correspond to the $T_{max}$ LLRs with the smallest magnitude. The associated bits are flipped in ascending order starting from the one with she smallest $|\alpha|$: thus, the construction of the list requires the $T_{max}$ LLRs to be not only selected, but also sorted. This operation results in substantial implementation cost, that is the main contributor to the additional hardware complexity of SC-Flip decoders with respect to SC decoders.

\begin{figure*}[t]
  \centering
   \scalebox{1}{\begin{tikzpicture}
  \pgfplotsset{
    label style = {font=\fontsize{9pt}{7.2}\selectfont},
    tick label style = {font=\fontsize{7pt}{7.2}\selectfont},
    yticklabel style={/pgf/number format/fixed}
  }

\begin{axis}[
	scale = 1,
    ybar interval,
    xlabel={Non-frozen indices}, xlabel style={yshift=0em},
    ylabel={Normalized $E_1$ Occurrence}, ylabel style={yshift=-0.75em},
    grid=both,
    xmin = -1,
    xmax = 201,
    ymin = 0,
    xtick={0,10,20,30,40,50,60,70,80,90,100,110,120,130,140,150,160,170,180},
    ymajorgrids=true,
    xmajorgrids=false,
    grid style=dashed,
    width=2*\columnwidth, height=7cm,
    thick,
    mark size=3,
    legend style={at={(0.27,0.93)},anchor=west},
    legend columns=2
]

\addplot[
    bar width=5pt,
    fill=red!60,opacity=0.8,
    thick,
]
table {
0	1
1	0.0845771144
2	0.5199004975
3	0
4	0.9353233831
5	0.3009950249
6	0.1206467662
7	0.0696517413
8	0
9	0.2089552239
10	0.0559701493
11	0.0074626866
12	0.0087064677
13	0
14	0.0099502488
15	0.0049751244
16	0.0012437811
17	0
18	0
19	0
20	0
21	0
22	0
23	0
24	0
25	0.8034825871
26	0.3519900498
27	0.1592039801
28	0
29	0.9676616915
30	0.2450248756
31	0.0684079602
32	0.0236318408
33	0.0111940299
34	0
35	0.118159204
36	0.013681592
37	0.0087064677
38	0.0024875622
39	0
40	0
41	0
42	0.0012437811
43	0
44	0
45	0
46	0
47	0
48	0
49	0
50	0
51	0
52	0
53	0
54	0
55	0
56	0
57	0.315920398
58	0.052238806
59	0.0049751244
60	0
61	0
62	0
63	0
64	0.0024875622
65	0
66	0
67	0
68	0
69	0
70	0
71	0
72	0
73	0
74	0
75	0
76	0
77	0
78	0
79	0
80	0
81	0
82	0
83	0
84	0
85	0
86	0
87	0
88	0
89	0
90	0.0049751244
91	0
92	0
93	0
94	0
95	0
96	0.0037313433
97	0
98	0
99	0.0012437811
100	0
101	0
102	0
103	0
104	0
105	0
106	0
107	0
108	0
109	0
110	0
111	0
112	0
113	0
114	0
115	0
116	0
117	0
118	0
119	0
120	0
121	0
122	0.0012437811
123	0
124	0
125	0
126	0
127	0
128	0
129	0
130	0
131	0
132	0
133	0
134	0
135	0
136	0
137	0
138	0
139	0
140	0
141	0
142	0
143	0
144	0
145	0
146	0
147	0
148	0
149	0
150	0
151	0
152	0
153	0
154	0
155	0
156	0
157	0
158	0
159	0
160	0
161	0
162	0
163	0
164	0
165	0
166	0
167	0
168	0
169	0
170	0
171	0
172	0
173	0
174	0
175	0
176	0
177	0

};
\addlegendentry{$E_b/N_0 = 2.5$ dB}

\addplot[
    ybar,
    bar width=2pt,
    fill=blue!80,opacity=0.6,
    thick,
    mark size=3,
]
table {
0	1
1	0.049382716
2	0.4074074074
3	0
4	0.7530864198
5	0.2427983539
6	0.0658436214
7	0.024691358
8	0
9	0.1316872428
10	0.0164609053
11	0.0082304527
12	0.0041152263
13	0
14	0
15	0
16	0
17	0
18	0
19	0
20	0
21	0
22	0
23	0
24	0
25	0.6666666667
26	0.2345679012
27	0.1234567901
28	0
29	0.7407407407
30	0.1358024691
31	0.0164609053
32	0.024691358
33	0.0041152263
34	0
35	0.037037037
36	0
37	0
38	0
39	0
40	0
41	0.0041152263
42	0
43	0
44	0
45	0
46	0
47	0
48	0
49	0
50	0
51	0
52	0
53	0
54	0
55	0
56	0
57	0.2098765432
58	0.012345679
59	0
60	0
61	0
62	0
63	0
64	0
65	0
66	0
67	0
68	0
69	0
70	0
71	0
72	0
73	0
74	0
75	0
76	0
77	0
78	0
79	0
80	0
81	0
82	0
83	0
84	0
85	0
86	0
87	0
88	0
89	0
90	0
91	0
92	0
93	0
94	0
95	0
96	0
97	0
98	0
99	0
100	0
101	0
102	0
103	0
104	0
105	0
106	0
107	0
108	0
109	0
110	0
111	0
112	0
113	0
114	0
115	0
116	0
117	0
118	0
119	0
120	0
121	0
122	0
123	0
124	0
125	0
126	0
127	0
128	0
129	0
130	0
131	0
132	0
133	0
134	0
135	0
136	0
137	0
138	0
139	0
140	0
141	0
142	0
143	0
144	0
145	0
146	0
147	0
148	0
149	0
150	0
151	0
152	0
153	0
154	0
155	0
156	0
157	0
158	0
159	0
160	0
161	0
162	0
163	0
164	0
165	0
166	0
167	0
168	0
169	0
170	0
171	0
172	0
173	0
174	0
175	0
176	0
177	0

};
\addlegendentry{$E_b/N_0 = 3.0$ dB}
\end{axis}
\end{tikzpicture}}
   \\
  \caption{Normalized distribution of $E_1$ occurrence for $PC(1024,170)$,~$C=8$, $5\times10^5$ simulated frames.}
  \label{fig:spikes}
\end{figure*}
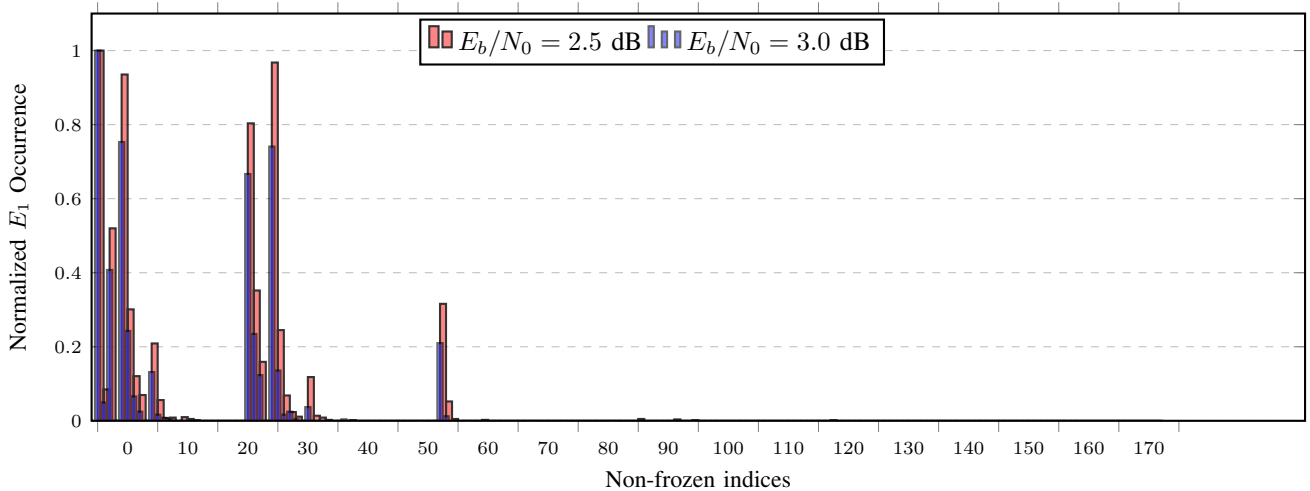

In Fig. \ref{fig:spikes} it is shown the normalized distribution of $E_1$ occurrence over the non-frozen bit indices for $PC(1024,170)$, with $C=8$, obtained simulating $5\times10^5$ frames, for two $E_b/N_0$ points. It can be noticed that bits have different probability of $E_1$, with a select few having a higher probability of $E_1$ than the remaining ones, and the majority of them not incurring any error. The height and position of the error distribution spikes can be used to determine, off-line, which bits to flip and in which order. In the provided example for $E_b/N_0 = 3.0$ dB, considering the non-frozen bits indexed from $0$ to $K+C-1$, the first flipped bit would be that at index $0$, corresponding to the highest $E_1$ occurrence, followed by those at index $29$, $4$, $25$, and so on, in descending probability of $E_1$ occurrence order, until $T_{max}$ attempts have been reached or the CRC check passes.
Determining which bits to flip off-line allows to avoid the LLR selection and sorting logic. Thus, the resulting SC-Flip decoder has an implementation cost negligibly different from that of an SC decoder.

\subsection{Enhanced Index Selection} \label{subsec:enhanced}

\begin{figure}
  \centering
   \scalebox{1}{\begin{tikzpicture}
  \pgfplotsset{
    label style = {font=\fontsize{9pt}{7.2}\selectfont},
    tick label style = {font=\fontsize{7pt}{7.2}\selectfont},
    yticklabel style={/pgf/number format/fixed}
  }

\begin{axis}[
	scale = 1,
    ybar interval,
    xlabel={Non-frozen indices}, xlabel style={yshift=0em},
    ylabel={Normalized average LLR magnitude}, ylabel style={yshift=-0.75em},
    grid=both,
    xmin = -1,
    xmax = 177,
    ymin = 0,
    xtick={0,20,40,60,80,100,120,140,160,180},
    ymajorgrids=true,
    xmajorgrids=false,
    grid style=dashed,
    width=1\columnwidth, height=7cm,
    thick,
    mark size=3,
    legend style={
      anchor={center},
      cells={anchor=west},
      column sep= 2mm,
      font=\fontsize{9pt}{7.2}\selectfont,
    },
    legend to name=LLRdist,
    legend columns=2,
]

\addplot[
    color=blue,
    thick,
    mark size=3,
]
table {
0	0.022062304
1	0.0310538868
2	0.0168963525
3	0.042218683
4	0.0150495956
5	0.0182988117
6	0.0209088959
7	0.0228052484
8	0.0546885625
9	0.0200868955
10	0.0236863344
11	0.0264554382
12	0.0284294438
13	0.0663354514
14	0.0283697805
15	0.0311617429
16	0.0331525829
17	0.0758841393
18	0.0345425813
19	0.0365542169
20	0.0826638059
21	0.0389624525
22	0.0875012378
23	0.090943225
24	0.1983503879
25	0.0154598119
26	0.0179211916
27	0.0197456676
28	0.048299967
29	0.0148915663
30	0.0188666447
31	0.0225058591
32	0.0252705067
33	0.027239561
34	0.0640286351
35	0.0210505859
36	0.0258841393
37	0.0299400066
38	0.0328762997
39	0.0349375309
40	0.0799839082
41	0.0325099026
42	0.0366849315
43	0.0396931837
44	0.0418361116
45	0.093932167
46	0.0418229906
47	0.0448114375
48	0.04694339
49	0.1042176927
50	0.0485154316
51	0.0506505199
52	0.1116042251
53	0.0532282555
54	0.1167610167
55	0.1204315894
56	0.2585723717
57	0.018790642
58	0.0243104473
59	0.0295934148
60	0.0338760522
61	0.036911289
62	0.0390882159
63	0.088675524
64	0.0324779667
65	0.0383613633
66	0.0428391649
67	0.0459905925
68	0.0482751279
69	0.1073559993
70	0.0457819772
71	0.0502858558
72	0.0534646806
73	0.0557615118
74	0.1224005611
75	0.055792375
76	0.0262348572
77	0.0589601419
78	0.0612049843
79	0.1333264565
80	0.0281748638
81	0.0628652418
82	0.0651118997
83	0.1410818617
84	0.0677994719
85	0.1464647632
86	0.1503540188
87	0.3192672058
88	0.0437894867
89	0.0501081862
90	0.0240061891
91	0.0547371678
92	0.0257045717
93	0.0580675029
94	0.060420779
95	0.1320812015
96	0.0241231226
97	0.0255578478
98	0.0579193761
99	0.0264888595
100	0.0279039445
101	0.0625363921
102	0.0295821093
103	0.0658288496
104	0.0681940089
105	0.1475565275
106	0.0292795016
107	0.0306751939
108	0.0681255158
109	0.0323493151
110	0.0713800132
111	0.0737115861
112	0.1585748473
113	0.0343163063
114	0.0753223304
115	0.0776368213
116	0.1663871926
117	0.0804442977
118	0.17192441
119	0.1758912362
120	0.3706040601
121	0.0266366562
122	0.0286561314
123	0.030101997
124	0.0671910381
125	0.0310608186
126	0.0324798647
127	0.071826374
128	0.0341944215
129	0.0751694174
130	0.0775384552
131	0.1663929691
132	0.0338805083
133	0.0353178742
134	0.0774942235
135	0.036986714
136	0.0807551576
137	0.0831085988
138	0.177475656
139	0.038981185
140	0.0847136491
141	0.0870382901
142	0.185256643
143	0.0898588876
144	0.1908202674
145	0.1947705892
146	0.4084007262
147	0.0371325301
148	0.0385751774
149	0.0840996864
150	0.0402930352
151	0.0874211916
152	0.0897978214
153	0.1908871101
154	0.0422465753
155	0.0913467569
156	0.0936920284
157	0.1986210596
158	0.0965217033
159	0.2041838587
160	0.2081828685
161	0.4351749464
162	0.044577323
163	0.0960934148
164	0.0984238323
165	0.2081944215
166	0.1012609341
167	0.2137316389
168	0.2176563789
169	0.4541912857
170	0.1045708863
171	0.2204282885
172	0.2243299224
173	0.4675482753
174	0.2291269186
175	0.4770745998
176	1
177	1
};

\addplot[
    color=red,
    thick,
    mark size=3,
]
table {
0	0.022062304
1	0.0310538868
2	0.0168963525
3	0
4	0.0150495956
5	0.0182988117
6	0.0209088959
7	0.0228052484
8	0
9	0.0200868955
10	0.0236863344
11	0.0264554382
12	0.0284294438
13	0
14	0.0283697805
15	0.0311617429
16	0.0331525829
17	0
18	0
19	0
20	0
21	0
22	0
23	0
24	0
25	0.0154598119
26	0.0179211916
27	0.0197456676
28	0
29	0.0148915663
30	0.0188666447
31	0.0225058591
32	0.0252705067
33	0.027239561
34	0
35	0.0210505859
36	0.0258841393
37	0.0299400066
38	0.0328762997
39	0
40	0
41	0
42	0.0366849315
43	0
44	0
45	0
46	0
47	0
48	0
49	0
50	0
51	0
52	0
53	0
54	0
55	0
56	0
57	0.018790642
58	0.0243104473
59	0.0295934148
60	0
61	0
62	0
63	0
64	0.0324779667
65	0
66	0
67	0
68	0
69	0
70	0
71	0
72	0
73	0
74	0
75	0
76	0
77	0
78	0
79	0
80	0
81	0
82	0
83	0
84	0
85	0
86	0
87	0
88	0
89	0
90	0.0240061891
91	0
92	0
93	0
94	0
95	0
96	0.0241231226
97	0
98	0
99	0.0264888595
100	0
101	0
102	0
103	0
104	0
105	0
106	0
107	0
108	0
109	0
110	0
111	0
112	0
113	0
114	0
115	0
116	0
117	0
118	0
119	0
120	0
121	0
122	0.0286561314
123	0
124	0
125	0
126	0
127	0
128	0
129	0
130	0
131	0
132	0
133	0
134	0
135	0
136	0
137	0
138	0
139	0
140	0
141	0
142	0
143	0
144	0
145	0
146	0
147	0
148	0
149	0
150	0
151	0
152	0
153	0
154	0
155	0
156	0
157	0
158	0
159	0
160	0
161	0
162	0
163	0
164	0
165	0
166	0
167	0
168	0
169	0
170	0
171	0
172	0
173	0
174	0
175	0
176	0
177	0
};

\end{axis}
\end{tikzpicture}}
  \\
  \vspace{2pt}
  \caption{Normalized average LLR magnitude at each non-frozen bit index for $PC(1024,170)$ at $E_b/N_0 = 2.5$ dB and $C=8$. Indices highlighted in red correspond to the non-zero occurrence of $E_1$ shown in Fig. \ref{fig:spikes}.}
  \label{fig:avgLLRs}
\end{figure}
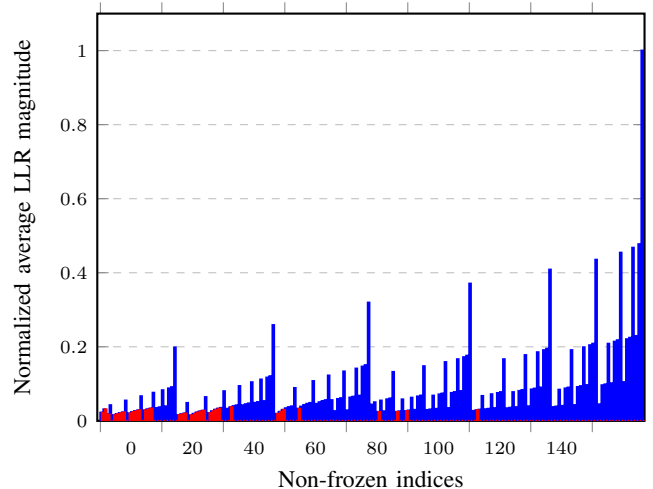

It is possible to use the error distribution information to increase the effectiveness of the standard LLR-based bit flipping selection scheme of SC-Flip. In its original formulation \cite{SCFlip14}, SC-Flip chooses which $T_{max}$ bits to flip among all the $K+C$ non-frozen bits in the code. However, from Fig. \ref{fig:spikes} we can see that the majority of information bits have very low error probability, that can be considered negligible depending on the target FER. Consequently, it is possible to restrict the LLR selection to the set of bit indices corresponding to high enough probability of $E_1$.

This can improve the error-correction performance by increasing the chance of the wrong estimation to be included within a limited number of $T_{max}$ indices, automatically excluding LLRs that have an averagely low magnitude, but correspond to indices with low error probability. Let us in fact observe Fig. \ref{fig:avgLLRs}, that shows the average LLR magnitude for each information bit index for $PC(1024,170)$, $C=8$, at $E_b/N_0 = 2.5$\,dB. It can be seen that the average LLR magnitude $|\alpha|$ can vary quite a lot depending on the bit. Highlighted in red are the LLRs whose indices correspond to the non-zero occurrences of $E_1$ shown in Fig. \ref{fig:spikes}: many LLRs of comparable or even smaller average magnitude are instead associated to indices with very low probability of $E_1$. These LLRs can feature among the $T_{max}$ with the smallest absolute value and be selected by the standard SC-Flip algorithm, but correspond to the erroneous bit in a negligible number of cases. On the other hand, their being taken in account can force the actual wrong estimation out of the $T_{max}$ flipped indices.   

The error distribution as shown in Fig. \ref{fig:spikes} depends on the number of simulated frames we are considering. With enough frames, every information bit index is associated with a non-zero $E_1$ occurrence: the zero error probability in Fig. \ref{fig:spikes} means that no errors were encountered over the $5 \times 10^5$ simulated frames. Given a target FER, it is possible to exclude from the considered indices all those associated to error probabilities lower than a certain threshold, in order to maximize the chances of identifying the wrong estimation within $T_{max}$ attempts. Moreover, to take in account the relative probability of $E_1$ occurrence, before sorting LLRs are scaled with respect to the the associated $E_1$ occurrence (see Fig. \ref{fig:spikes}).

\section{Simulation Results} \label{sec:sims}

\begin{figure}
  \centering
   \scalebox{1}{\begin{tikzpicture}[spy using outlines=
	{rectangle, magnification=2, connect spies}]
  \pgfplotsset{
    label style = {font=\fontsize{9pt}{7.2}\selectfont},
    tick label style = {font=\fontsize{7pt}{7.2}\selectfont}
  }

\begin{axis}[
	scale = 1,
    ymode=log,
    xlabel={$E_b/N_0$ [\text{dB}]}, xlabel style={yshift=0.4em},
    ylabel={FER}, ylabel style={yshift=-0.75em},
    grid=both,
    ymajorgrids=true,
    xmajorgrids=true,
    grid style=dashed,
    width=1\columnwidth, height=7cm,
    thick,
    mark size=3,
    legend style={
      anchor={center},
      cells={anchor=west},
      column sep= 2mm,
      font=\fontsize{7pt}{7.2}\selectfont,
    },
    legend to name=ECP-R0.16,
    legend columns=2,
]

\addplot[
    color=black,
    dashed,
    mark=x,
    thick,
    mark size=3,
]
table {
1.0 6.23980e-01
1.5 3.75280e-01
2.0 1.66320e-01
2.5 5.23600e-02
3.0 1.09200e-02
3.5 2.18000e-03
4.0	1.43912e-04
};
\addlegendentry{SC}

\addplot[
    color=red,
    mark=triangle,
    thick,
    mark size=3,
]
table {
1.00	4.81320e-01
1.50	2.34650e-01
2.00	7.87200e-02
2.50	1.75900e-02
3.00	2.69000e-03
3.50	3.40000e-04
4.00	2.48243e-05
};
\addlegendentry{SC-Flip}

\addplot[
    color=green!60!black,
    mark=o,
    thick,
    mark size=3,
]
table {
1.00	4.51070e-1
1.50	2.05370e-1
2.00	6.45700e-2
2.50	1.34200e-2
3.00	2.05000e-3
3.50	1.92172e-4
4.00	9.05877e-6
};
\addlegendentry{SC-Flip - FIS}

\addplot[
    color=blue,
    mark=diamond,
    thick,
    mark size=3,
]
table {
1.00	4.05650e-1
1.50	1.63560e-1
2.00	4.33000e-2
2.50	6.66000e-3
3.00	6.40000e-4
3.50	3.44288e-5
};
\addlegendentry{SC-Flip - EIS}

\addplot[
    color=black,
    mark=+,
    thick,
    mark size=3,
]
table {
1.0 3.53100e-01
1.5 1.45200e-01
2.0 3.93000e-02
2.5 7.50000e-03
3.0 5.99003e-04
3.5 4.08481e-05
};
\addlegendentry{SC-List ($L=2$)}


\addplot[
    color=orange,
    mark=square,
    thick,
    mark size=3,
]
table {
1.0	3.54180e-1
1.5	1.37380e-1
2.0	3.29700e-2
2.5	4.70000e-3
3.0	3.30000e-4
3.5	1.27870e-5
};
\addlegendentry{SC-Oracle}

\coordinate (spypoint) at (axis cs:3.0,8e-4);
\coordinate (magnifyglass) at (axis cs:1.45,8e-4);

\end{axis}
\spy [magenta, height=2.6cm, width=2.2cm] on (spypoint)
   in node[fill=white] at (magnifyglass);
\end{tikzpicture}}
   \ref{ECP-R0.16}
  \\
  \vspace{2pt}
  \caption{FER curves for for $PC(1024,170)$,~$C = 8$,~$T_{max} = 10$ with different decoding approaches.}
  \label{fig:curves016}
\end{figure}
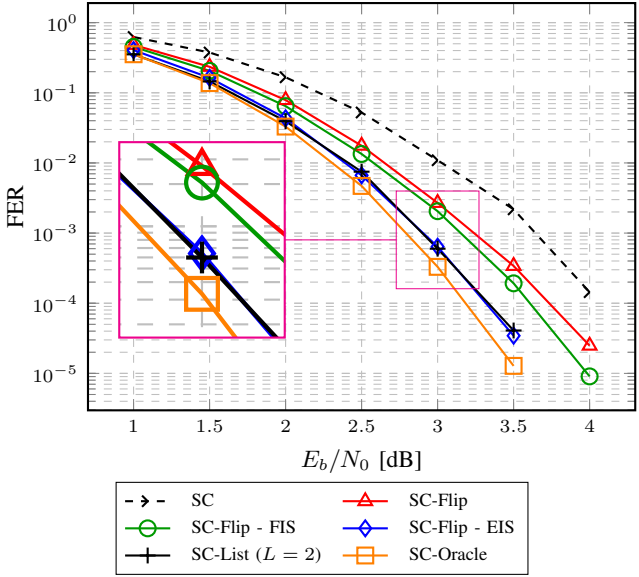

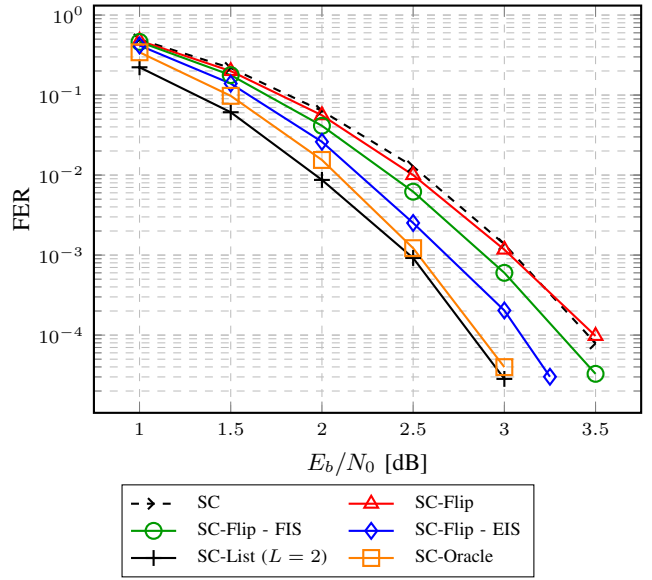
\begin{figure}
  \centering
   \scalebox{1}{\begin{tikzpicture}[spy using outlines=
	{rectangle, magnification=2, connect spies}]
  \pgfplotsset{
    label style = {font=\fontsize{9pt}{7.2}\selectfont},
    tick label style = {font=\fontsize{7pt}{7.2}\selectfont}
  }

\begin{axis}[
	scale = 1,
    ymode=log,
    xlabel={$E_b/N_0$ [\text{dB}]}, xlabel style={yshift=0.4em},
    ylabel={FER}, ylabel style={yshift=-0.75em},
    grid=both,
    ymajorgrids=true,
    xmajorgrids=true,
    grid style=dashed,
    width=1\columnwidth, height=7cm,
    thick,
    mark size=3,
    legend style={
      anchor={center},
      cells={anchor=west},
      column sep= 2mm,
      font=\fontsize{7pt}{7.2}\selectfont,
    },
    legend to name=ECP-R0.25,
    legend columns=2,
]

\addplot[
    color=black,
    dashed,
    mark=x,
    thick,
    mark size=3,
]
table {
1.0 4.94580e-01
1.5 2.17820e-01
2.0 6.33400e-02
2.5 1.28000e-02
3.0 1.36763e-03
3.5 7.89148e-05
};
\addlegendentry{SC}

\addplot[
    color=red,
    mark=triangle,
    thick,
    mark size=3,
]
table {
1.0	4.78220e-01
1.5 1.99620e-01
2.0 5.59000e-02
2.5 1.00000e-02
3.0 1.17000e-03
3.5 9.75775e-05
};
\addlegendentry{SC-Flip}

\addplot[
    color=green!60!black,
    mark=o,
    thick,
    mark size=3,
]
table {
1.0	4.61890e-01
1.5	1.77630e-01
2.0	4.13300e-02
2.5	6.20000e-03
3.0	6.00000e-04
3.5	3.27964e-05
};
\addlegendentry{SC-Flip - FIS}

\addplot[
    color=blue,
    mark=diamond,
    thick,
    mark size=3,
]
table {
1.0	4.16410e-01
1.5	1.39250e-01
2.0	2.60600e-02
2.5	2.52000e-03
3.0	2.03132e-04
3.25	3.01738e-05
};
\addlegendentry{SC-Flip - EIS}

\addplot[
    color=black,
    mark=+,
    thick,
    mark size=3,
]
table {
1.0 2.22400e-01
1.5 6.10000e-02
2.0 8.70000e-03
2.5 9.20334e-04
3.0 2.83470e-05
};
\addlegendentry{SC-List ($L=2$)}

\addplot[
    color=orange,
    mark=square,
    thick,
    mark size=3,
]
table {
1.0	3.42580e-01
1.5 9.78000e-02
2.0 1.54500e-02
2.5 1.22000e-03
3.0 3.97831e-05
};
\addlegendentry{SC-Oracle}

\end{axis}
\end{tikzpicture}}
   \ref{ECP-R0.25}
  \\
  \vspace{2pt}
  \caption{FER curves $PC(1024,256)$,~$C = 8$,~$T_{max} = 10$ with different decoding approaches.}
  \label{fig:curves025}
\end{figure}

The error-correction performance of the proposed methods has been evaluated and compared against the state of the art.
Fig. \ref{fig:curves016} and Fig. \ref{fig:curves025} show the FER for $PC(1024,170)$ and $PC(1024,256)$ respectively, with $C=8$. Both codes have been constructed targeting a signal-to-noise ratio of $2.5$\,dB, and simulations have been run on AWGN channel and BPSK modulation. Along with standard SC, curves are plot for SC-Flip \cite{SCFlip14}, SC-List \cite{TalList},  SC-Oracle, and SC-Flip with the proposed fixed index selection (FIS) and enhanced index selection (EIS). The maximum number of iterations has been set to $T_{max}=10$ for all flip-based algorithms. The SC-Oracle curves have been obtained considering K+C non-frozen bits.

The FER of SC is the highest among all considered algorithms, since all incarnations of SC-Flip substantially improve on its error correction performance. One exception to this is SC-Flip with $PC(1024,256)$; its performance matches that of SC, due to the degradation caused by errors on the additional non-frozen CRC bits. On the other hand, the SC-Oracle curve represents the case in which all $E_1$ occurrences are successfully corrected, and thus can be interpreted as the lower bound of flip-based FER. The SC-Flip decoding algorithm improves the error-correction performance of SC, but in both Fig. \ref{fig:curves016} and Fig. \ref{fig:curves025} a substantial gap from the SC-Oracle bound can be observed, consisting of $0.55$\,dB for $PC(1024,170)$ and of $0.63$\,dB for $PC(1024,256)$ at FER=$10^{-4}$. 

SC-Flip with FIS is shown to perform slightly better than standard SC-Flip for $PC(1024,170)$ in Fig. \ref{fig:curves016}, with a gain of $0.12$\,dB at FER=$10^{-4}$. In case of $PC(1024,256)$, the gain of the FIS criterion compared to baseline SC-Flip is $0.19$\,dB. The $T_{max}=10$ indices considered by the method are those corresponding to the highest probabilities of $E_1$. This technique allows to substantially reduce the implementation complexity of an SC-Flip decoder. The second column of Table \ref{tab:complex} reports the number and type of operations required by a standard SC-Flip decoder, assuming a semi-parallel structure similar to \cite{Raymond_TSP14}, and omitting the memory requirements. The complexity is divided among the three main operations that an SC-Flip decoder has to perform: F (\ref{eqn:alphaleft}), G (\ref{eqn:alpharight}), C (\ref{eqn:beta}), and the sorting and selection of the $T_{max}$ indices to flip. Basic modules as comparators, XORs, multiplexers, adders and registers are used to approximate the implementation complexity; the number of parallel processing elements $Pe$, along with the LLR quantization bits $Q$, also affect the total cost. It can be seen that FIS allows to completely avoid the logic cost of the sorter function. To get a sense of its impact on the total logic complexity, we consider the XOR block having a cost of $1$, the multiplexer a cost of $3$, the adder and the comparator a cost of $5$ and the register a cost of $4$. Supposing an architecture with $Pe=32$, $Q=6$ and $T_{max}=10$, the sorter accounts for $24.6\%$ of the total logic complexity.

\begin{table}
\centering
\caption{Implementation logic cost breakdown for SC-Flip and SC-Flip~-~FIS.}
\label{tab:complex}
\setlength{\extrarowheight}{1.1pt}
\begin{tabular}{l | c | c}
\hline
 & SC-Flip & SC-Flip~-~FIS \\
\hline
\multirow{3}{*}{F} & $P_e \times Q \times \text{Comparator}$ & $P_e \times Q \times \text{Comparator}$ \\
& $2 \times P_e \times \text{XOR}$ & $2 \times P_e \times \text{XOR}$ \\
& $P_e \times Q \times \text{MUX}$ & $P_e \times Q \times \text{MUX}$ \\
\hline
\multirow{2}{*}{G} & $P_e \times Q \times \text{Sum}$ & $P_e \times Q \times \text{Sum}$ \\
& $P_e \times \text{MUX}$ & $P_e \times \text{MUX}$ \\
\hline
C & $P_e \times \text{MUX}$ & $P_e \times \text{MUX}$ \\
\hline
\multirow{3}{*}{Sorter} &  $T_{max} \times Q \times \text{D-FF}$ &\\
 & $T_{max} \times Q \times \text{Comparator}$ & N/A\\
& $2 \times T_{max} \times Q \times \text{MUX}$& \\
\hline
\end{tabular}
\end{table}

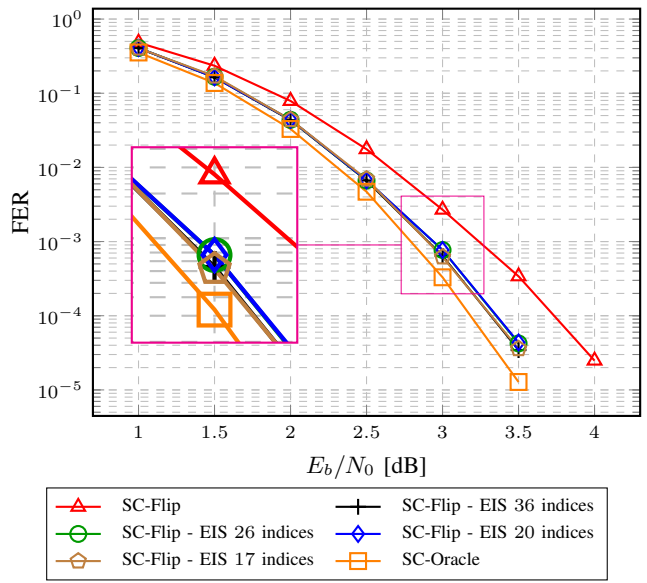
\begin{figure}
  \centering
   \scalebox{1}{\begin{tikzpicture}[spy using outlines=
	{rectangle, magnification=2, connect spies}]
  \pgfplotsset{
    label style = {font=\fontsize{9pt}{7.2}\selectfont},
    tick label style = {font=\fontsize{7pt}{7.2}\selectfont}
  }

\begin{axis}[
	scale = 1,
    ymode=log,
    xlabel={$E_b/N_0$ [\text{dB}]}, xlabel style={yshift=0.4em},
    ylabel={FER}, ylabel style={yshift=-0.75em},
    grid=both,
    ymajorgrids=true,
    xmajorgrids=true,
    grid style=dashed,
    width=1\columnwidth, height=7cm,
    thick,
    mark size=3,
    legend style={
      anchor={center},
      cells={anchor=west},
      column sep= 2mm,
      font=\fontsize{7pt}{7.2}\selectfont,
    },
    legend to name=ECP-thresholded-R0.16,
    legend columns=2,
]

\addplot[
    color=red,
    mark=triangle,
    thick,
    mark size=3,
]
table {
1.00	4.81320e-01
1.50	2.34650e-01
2.00	7.87200e-02
2.50	1.75900e-02
3.00	2.69000e-03
3.50	3.40000e-04
4.00	2.48243e-05
};
\addlegendentry{SC-Flip}

\addplot[
    color=black,
    mark=+,
    thick,
    mark size=3,
]
table {
1.00	4.05650e-1
1.50	1.63560e-1
2.00	4.33000e-2
2.50	6.66000e-3
3.00	6.40000e-4
3.50	3.44288e-5
};
\addlegendentry{SC-Flip - EIS $36$ indices}

\addplot[
    color=green!60!black,
    mark=o,
    thick,
    mark size=3,
]
table {
1.00	4.06810e-01
1.50	1.66610e-01
2.00	4.38100e-02
2.50	6.66000e-03
3.00	7.70000e-04
3.50	4.23330e-05
};
\addlegendentry{SC-Flip - EIS $26$ indices}

\addplot[
    color=blue,
    mark=diamond,
    thick,
    mark size=3,
]
table {
1.00	4.07490e-1
1.50	1.64190e-1
2.00	4.38900e-2
2.50	6.72000e-3
3.00	7.70000e-4
3.50	4.35640e-5
};
\addlegendentry{SC-Flip - EIS $20$ indices}

\addplot[
    color=brown,
    mark=pentagon,
    thick,
    mark size=3,
]
table {
1.00	4.05370e-1
1.50	1.69980e-1
2.00	4.40200e-2
2.50	6.95000e-3
3.00	6.20000e-4
3.50	3.59128e-5
};
\addlegendentry{SC-Flip - EIS $17$ indices}

\addplot[
    color=orange,
    mark=square,
    thick,
    mark size=3,
]
table {
1.0	3.54180e-1
1.5	1.37380e-1
2.0	3.29700e-2
2.5	4.70000e-3
3.0	3.30000e-4
3.5	1.27870e-5
};
\addlegendentry{SC-Oracle}

\coordinate (spypoint) at (axis cs:3.0,9e-4);
\coordinate (magnifyglass) at (axis cs:1.5,9e-4);

\end{axis}
\spy [magenta, height=2.6cm, width=2.2cm] on (spypoint)
   in node[fill=white] at (magnifyglass);
\end{tikzpicture}}
   \ref{ECP-thresholded-R0.16}
  \\
  \vspace{2pt}
  \caption{FER curves for $PC(1024,170),~C = 8,~T_{max} = 10$ with different decoding approaches and number of indices considered by SC-Flip - EIS.}
  \label{fig:thresholded016}
\end{figure}

Compared to SC-Flip, the EIS approach allows for $0.4$\,dB and $0.42$\,dB gain for $PC(1024,170)$ and $PC(1024,256)$ respectively. In Fig. \ref{fig:curves016}, a total of $26$ indices are considered, while $45$ are taken in account for the higher rate case portrayed in Fig. \ref{fig:curves025}. The FER of SC-Flip with EIS matches to that of SC-List with $L=2$. In the case of $PC(1024,256)$, the error-correction performance of the list decoder outperforms SC-Oracle decoder.
It was shown in \cite{eMBBcode-Asilomar17} that while an outer CRC code improves the error-correction performance of SC-List decoding significantly at medium to high code rates, at low code rates the CRC is detrimental. Thus, for a fair comparison, we did not consider any CRC in the SC-List curves.
The indices used by EIS have been found simulating a total of $1\times10^6$ frames, and selecting those with the higher probability of $E_1$. The impact of different number of considered indices in EIS is shown in Fig. \ref{fig:thresholded016} for $PC(1024,170)$: it can be seen that decreasing it tends to degrade the error-correction performance, while increasing it does not bring substantial advantage. This remains true as long as the number of indices considered by EIS is substantially lower than $K+C$, at which point the FER will degrade to the same obtained by standard SC-Flip.

As the code rate increases, the error correction performance of both FIS and EIS starts degrading if $T_{max}$ is not increased. This is due to the fact that as the number of information bits in the code rises, the distribution of $E_1$ occurrence spreads over more bit indices. This results in a higher number of indices that need to be considered by the proposed method: a higher $T_{max}$ is thus needed to maintain the error-correction performance.

 \begin{figure}
   \centering
    \scalebox{1}{\begin{tikzpicture}
  \pgfplotsset{
    label style = {font=\fontsize{9pt}{7.2}\selectfont},
    tick label style = {font=\fontsize{7pt}{7.2}\selectfont}
  }

\begin{axis}[
	scale = 1,
    xlabel={$E_b/N_0$ [\text{dB}]}, xlabel style={yshift=0.4em},
    ylabel={Average number of iterations}, ylabel style={yshift=-0.75em},
    grid=both,
    ymajorgrids=true,
    xmajorgrids=true,
    grid style=dashed,
    width=1\columnwidth, height=7cm,
    thick,
    mark size=3,
    legend style={
      anchor={center},
      cells={anchor=west},
      column sep= 2mm,
      font=\fontsize{7pt}{7.2}\selectfont,
    },
    legend to name=iter-thresholded-R0.16,
    legend columns=2,
]

\addplot[
    color=black,
    thick,
    mark size=3,
]
table {
1.0 1
3.5 1
};
\addlegendentry{SC}

\addplot[
    color=blue,
    mark=o,
    thick,
    mark size=3,
]
table {
1.0	2.22
1.5	1.68
2.0	1.27
2.5	1.07
3.0	1.01
3.5	1.00
};
\addlegendentry{SC-Flip}

\addplot[
    color=red,
    mark=x,
    thick,
    mark size=3,
]
table {
1.0	3.20
1.5	2.20
2.0	1.49
2.5	1.14
3.0	1.03
3.5	1.00
};
\addlegendentry{SC-Flip - FIS}

\addplot[
    color=green!50!black,
    mark=triangle,
    thick,
    mark size=3,
]
table {
1.0	2.92
1.5	1.93
2.0	1.31
2.5	1.07
3.0	1.01
3.5	1.00
};
\addlegendentry{SC-Flip - EIS 26 indices}

\addplot[
    color=brown,
    mark=triangle,
    thick,
    mark size=3,
]
table {
1.0	2.92
1.5	1.92
2.0	1.31
2.5	1.07
3.0	1.01
3.5	1.00
};
\addlegendentry{SC-Flip - EIS 20 indices}

\addplot[
    color=orange,
    mark=pentagon,
    thick,
    mark size=3,
]
table {
1.0	2.93
1.5	1.93
2.0	1.31
2.5	1.07
3.0	1.01
3.5	1.00
};
\addlegendentry{SC-Flip - EIS 17 indices}

\end{axis}
\end{tikzpicture}}
    \ref{iter-thresholded-R0.16}
   \\
   \vspace{2pt}
   \caption{Average number of iterations for $PC(1024,170)$,~$C = 8$,~$T_{max} = 10$ with different decoding approaches.}
   \label{fig:avgiter}
 \end{figure}
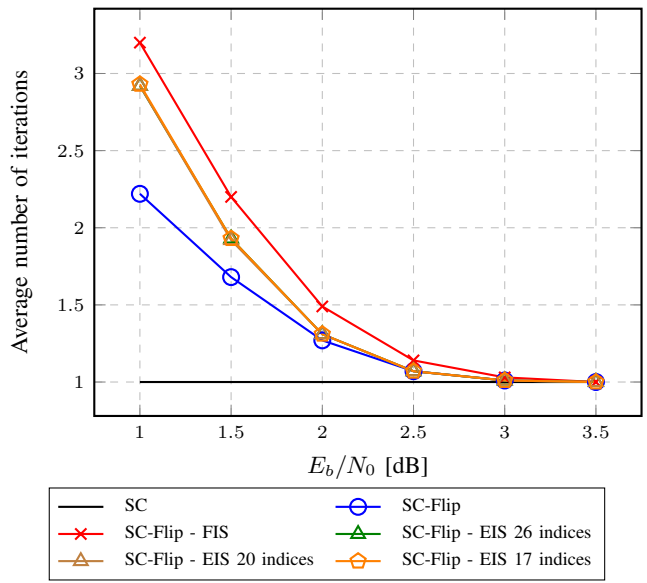
 
Fig. \ref{fig:avgiter} depicts the average number of iterations for different decoding approaches, where an iteration is defined as an application of the SC algorithm over the $N$ codeword bits. It can be observed that at low $E_b/N_0$ points, SC-Flip with the FIS technique has up to $44\%$ higher average iterations with respect to standard SC-Flip \cite{SCFlip14}, while the use of EIS entails an increment of up to $30\%$. Both EIS and FIS reduce the number of cases in which SC-Flip reaches the $T_{max}$ attempts, lowering the average number of iterations. For example, the SC-Flip decoding algorithm with $PC(1024,170)$ at $E_b/N_0 = 1.0$\,dB is not able to find the correct codeword $45\%$ of the time, with $T_{max}=10$ and $C=8$. With the EIS criterion, this failure rate drops by $12\%$. However, this positive contribution is offset by the position of the indices with high probability of $E_1$ occurrence, as shown in Fig. \ref{fig:spikes}-\ref{fig:avgLLRs}. The majority of these indices are in fact found towards the beginning of the codeword: thus, each additional decoding attempt requires almost a full iteration. On the contrary, in standard SC-Flip there are no constraints on which indices to consider, and at low $E_b/N_0$, the channel noise causes most LLRs to have very low magnitude. Consequently, the $T_{max}$ indices can be spread out over the whole codeword, leading to standard SC-Flip having a lower average number of iterations. At higher $E_b/N_0$, where the FER is of interest, the average number of iterations quickly converges to $1$ for all methods. 

\section{Conclusion}\label{sec:concl}

In this work, we have proposed two techniques to reduce the implementation complexity and reduce the FER of the SC-Flip decoding algorithm. They are based on the identification of the bit indices where a channel-induced error is most likely to occur. The proposed techniques have shown significant improvement over SC-Flip when applied to low-rate polar codes. The fixed index selection method has shown an estimated $24.6\%$ implementation complexity reduction at no cost in error-correction performance, while the enhanced index selection has been observed to bring up to $0.42$\,dB gain in FER.


\end{document}